\documentclass{article}
\usepackage{spconf,amsmath,graphicx}

\title{FURCANET: AN END-TO-END DEEP GATED CONVOLUTIONAL, LONG SHORT-TERM MEMORY,
DEEP NEURAL NETWORKS FOR SINGLE CHANNEL SPEECH SEPARATION}

\name{Ziqiang Shi$^1$, Huibin Lin$^1$, Liu Liu$^1$, Rujie Liu$^1$, Shoji Hayakawa$^2$, Shouji Harada$^2$, Jiqing Han$^3$}
\address{Fujitsu Research and Development Center, Beijing, China$^1$\\
Fujitsu Laboratories Ltd. Kawasaki, Japan$^2$ \\
Harbin Institute of Technology, Harbin, China$^3$
}

\begin{document}
%
\maketitle
\begin{abstract}
Deep gated convolutional networks have been proved to be very effective in single channel speech separation. However current state-of-the-art framework often considers training the gated convolutional networks in time-frequency (TF) domain. Such an approach will result in limited perceptual score, such as signal-to-distortion ratio (SDR) upper bound of separated utterances and also fail to exploit an end-to-end framework. In this paper we present an integrated simple and effective end-to-end approach to monaural speech separation, which consists of deep gated convolutional neural networks (GCNN) that takes the mixed utterance of two speakers and maps it to two separated utterances, where each utterance contains only one speaker's voice. In addition long short-term memory (LSTM) is employed for long term temporal modeling. For the objective, we propose to train the network by directly optimizing utterance level SDR in a permutation invariant training (PIT) style. Our experiments on the public WSJ0-2mix data corpus demonstrate that this new scheme can produce more discriminative separated utterances and leading to performance improvement on the speaker separation task.
\end{abstract}
\section{Introduction}
\label{sec:intro}

Multi-talker monaural speech separation has a vast range of applications. For example, a home environment or a conference environment in which many people talk, the human auditory system can easily track and follow a target speaker's voice from the multi-talker's mixed voice. In this case, if automatic speech recognition and speaker recognition are to be performed, a clean speech signal of the target speaker needs to be separated from the mixed speech to complete the subsequent recognition work. Thus it is a problem that must be solved in order to achieve satisfactory performance in speech or speaker recognition tasks. There are two difficulties in this problem, the first is that since we don't have any priori information of the user, a truly practical system must be speaker-independent. The second difficulty is that there is no way to use the beamforming algorithm for a single microphone signal. Many traditional methods, such as computational auditory scene analysis (CASA)~\cite{wang2006computational,shao2006model,hu2013unsupervised}, Non-negative matrix factorization (NMF)~\cite{smaragdis2007convolutive, le2015sparse}, and probabilistic models ~\cite{virtanen2006speech}, do not solve these two difficulties well.

\begin{figure*}[htb]
\centering
\includegraphics[width=0.9\textwidth]{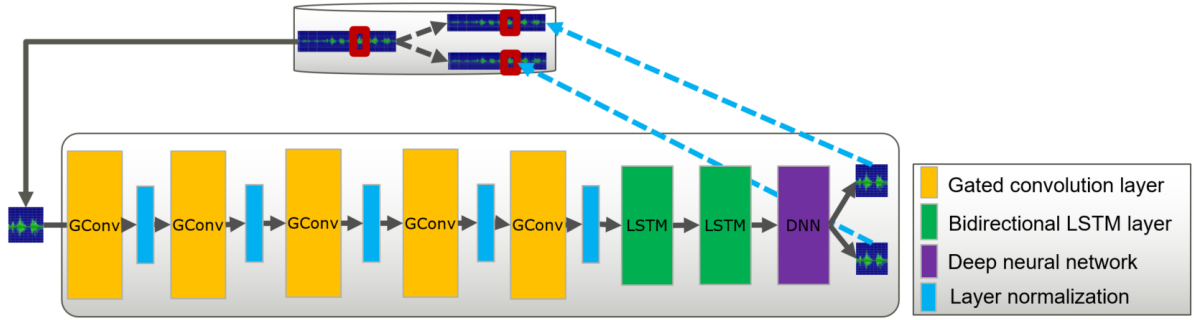}
\caption{Architecture of FurcaNet.}
\label{fig:furcanet}
\end{figure*}

More recently, a large number of techniques based on deep learning are proposed for this task. These methods can be briefly grouped into three categories. The first category is based on deep clustering (DPCL)~\cite{hershey2016deep,isik2016single}, which maps the time-frequency (TF) points of the spectrogram into the embedding vectors, then these embedding vectors are clustered into several classes corresponding to different speakers, and finally these clusters are used as masks to inversely transform the spectrogram to the separated clean voices; the second is the permutation invariant training (PIT)~\cite{kolbaek2017multitalker,yu2017permutation}, which solves the label permutation problem by minimizing the lowest error output among all possible permutations for N mixing sources assignment; the third category is end-to-end speech separation in time-domain~\cite{luo2017tasnet,venkataramani2017adaptive,luo2018tasnet}, which is a natural way to overcome the obstacles of the upper bound source-to-distortion ratio improvement (SDRi) in short-time Fourier transform (STFT) mask estimation based methods and real-time processing requirements in actual use.

This paper is based on the end-to-end method~\cite{luo2017tasnet,venkataramani2017adaptive,luo2018tasnet}, which has achieved better results than DPCL based or PIT based approaches. Since most DPCL and PIT based methods use STFT as front-end. Specifically, the mixed speech signal is first transformed from one-dimensional signal in time domain to two-dimensional spectrum signal in TF domain, and then the mixed spectrum is separated to result in spectrums corresponding to different source speeches by a deep clustering method, and finally the cleaned source speech signal can be restored by an inverse STFT on each spectrum. This framework has several limitations. Firstly, it is unclear whether the STFT  is the optimal (even assume the parameters it depends on are optimal, such as size and overlap of audio frames, window type and so on) transformation of the signal for speech separation. Secondly, most STFT based methods often assumed that the phase of the separated signal to be equal to the mixture phase, which is generally incorrect and imposes an obvious upper bound on separation performance by using the ideal masks. As an approach to overcome the above problems, several
speech separation models were recently proposed that operate
directly on time-domain speech signals~\cite{luo2017tasnet,venkataramani2017adaptive}. Inspired by these first results, we propose FurcaNet\footnotemark[1], a fully end-to-end time-domain separation system, based on deep gated convolutional neural network (GCNN)~\cite{dauphin2016language}, bidirectional long short-term memory (BiLSTM), deep neural network (DNN), which has showed promising performance on both a clean and noisy Voice Search tasks~\cite{sainath2015learning}.

\footnotetext[1]{Furca is Latin for ``fork'', and we use this word to mean the speech is split into two streams by our network like water.}

The remainder of this paper is organized as follows:  section 2 introduces monaural
speech separation, describe our proposed FurcaNet and the separation algorithm in detail. The experimental
setup and results are presented in Section 3. We conclude this
paper in Section 4.

\section{THE FURCANET MODEL}
\label{sec:furcanet}

The proposed end-to-end deep learning approach consists of two main components: one is the FurcaNet pipeline, which consists of GCNN, BiLSTM and DNN; and the other is the perceptual loss function.

In this section, we first review the formal definition of the monaural speech separation task and the GCNN architecture.
The details of the FurcaNet structure we investigated will be introduced.
Finally the perceptual metric as a loss function is introduced.

\subsection{Monaural speech separation}

The goal of monaural speech separation is to estimate the individual target signals from a linearly mixed single-microphone signal, in which the target signals overlap in the TF domain.
Let $x_i(t),i=1,..,S$ denote the $S$ target speech signals and  $y(t)$ denotes the
mixed speech respectively. If we assume the target signals are linearly mixed, which can be represented as:
\begin{equation*}
y(t)=\sum_{i=1}^{S}x_i(t),
\end{equation*}
then monaural speech separation aims at estimating individual target signals from
given mixed speech $y(t)$. In this work it is assumed that the number of target signals is known.

In this work, we propose an end-to-end deep learning approach to separate the mixed utterance. The input of the FurcaNet is a mixed utterance $y(t)$, and the output of the network are the separated utterances, ideally it is best to be exactly the same $x_i(t),i=1,..,S$. In order to do this, firstly the mixed speech is framed.
Then each frame of the mixed utterance $y(t)$ is directly as raw wave forward propagated through the FurcaNet, and the output activations are the  separated frames, each frame is corresponding only one speaker. Finally the separated frames are concatenate together to form the output utterances.

\subsection{Network architecture}

The proposed FurcaNet model is similar to~\cite{sainath2015learning}, but with fine
adjustment. The FurcaNet separation system comprises GCNN, BiLSTM and DNN, and the structure is illustrated in Fig.~\ref{fig:furcanet}. A deep GCNN proposed in~\cite{dauphin2016language} is adopted here to build the front-end. GCNN is implemented by stacking multiple 1D gated convolutional (GConv) layers on top of each other.

\begin{figure}[htb]
\centering
\includegraphics[width=0.45\textwidth]{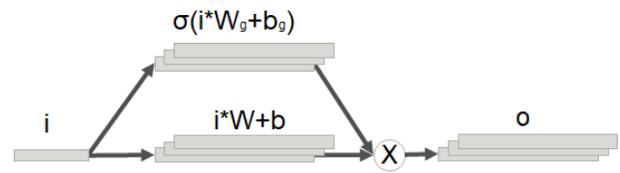}
\caption{Architecture of a 1D GConv layer.}
\label{fig:gcnns}
\end{figure}

Fig.~\ref{fig:gcnns} shows the structure of a 1D GConv layer.
The main difference between a GConv layer and a plain
convolutional layer is a gated linear unit (GLU)~\cite{dauphin2016language}, namely the
gates $\sigma(i*W_g+b_g)$ of Eq.~(\ref{eq:gcnns}) is used as a nonlinear control function
instead of tanh activation or regular rectified linear units (ReLUs)~\cite{dauphin2016language}:
\begin{equation}
o=(i*W+b)\otimes \sigma(i*W_g+b_g),
\label{eq:gcnns}
\end{equation}
where $i$ and $o$ are the input and output, $W$, $b$, $W_g$, and $b_g$ are learned parameters, $\sigma$ is the sigmoid function and $\otimes$ is the element-wise product between vectors or matrices.
Similar to LSTMs, GLUs play the role of controlling the information passed on
in the hierarchy.
This special gating mechanism allows us to effectively capture long-range context dependencies by deepening layers without encountering the problem of vanishing gradient. In order to stabilize the training and also reduce the training time,
a layer normalization(LNrom) operator~\cite{ba2016layer} was added behind each GConv layer.

In order to capture long-term contextual dependencies,
BiLSTM is applied to replace the bottleneck architectures or dilated convolutional networks~\cite{li2018deep}.
BiLSTM is a natural choice for modeling long-term time series data since the recurrent connection architectures allow the network to make prediction with the entire input time series. After the gated convolution, we pass the GCNN output to BiLSTM layers. Finally we pass the output of the BiLSTM to one fully connected DNN layer. The DNN layer maps the signal further to a more separable space.
The FurcaNet incorporates the GConv, BiLSTM, and DNN layers into a unified framework, combines the advantages of different layers. All the layers are trained jointly. During training we need to provide the correct reference $x_i(t),i=1,..,S$ to the corresponding output layer for supervision.

\subsection{Perceptual metric: Utterance-level SDR objective}
\label{sec:loss}

Since the loss function of many STFT-based methods is not directly applicable to waveform-based end-to-end speech separation, perceptual metric based loss function is tried in this work. The perception of speech is greatly affected by distortion~\cite{yang1998performance,assmann2004perception}.
Generally in order to evaluate the performance of speech separation, the BSS\_Eval metrics signal-to-distortion ratio (SDR), signal-to-Interference
ratio (SIR), signal-to-artifact ratio (SAR)~\cite{fevotte2005bss,vincent2006performance},
and short-time objective intelligibility (STOI)~\cite{taal2010short}
have been often employed. In this work we directly use SDR, which is most commonly used metrics to evaluate the performance of source separation, as the training objective. SDR measures the amount of distortion introduced by the output signal and define it as the ratio between the energy of the clean signal and the energy of the distortion.

SDR captures the overall separation quality of the algorithm. There is a subtle problem here. We first concatenate the outputs of FurcaNet into a complete utterance and then compare with the input full utterance to calculate the SDR in the utterance level instead of calculating the SDR for one frame at a time. These two methods are very different in ways and performance. If we denote the output of the network by $s$, which should ideally be equal to the target source $x$, then SDR can be given as~\cite{fevotte2005bss,vincent2006performance}
\begin{eqnarray*}
 \tilde{x}&=&\frac{\langle x , s \rangle}{\langle x , x \rangle} x, \\ e&=&\tilde{x}-s,\\ \text{SDR} &=& 10*\text{log}_{10}\frac{\langle \tilde{x} , \tilde{x} \rangle}{\langle e , e \rangle}.
\end{eqnarray*}
Then our target is to maximize SDR or minimize the negative SDR as loss function respect to the $s$.

In order to solve tracing and permutation problem, the PIT training criteria~\cite{kolbaek2017multitalker,yu2017permutation} is employed in this work. We calculate the SDRs for all the permutations, pick the maximum one, and take the negative as the loss. It is called the uSDR loss in this work.

\section{EXPERIMENTS}
\label{sec:experiments}

\subsection{Dataset and neural network}
\label{ssec:dataset}

We evaluated our system on two-speaker speech separation problem using WSJ0-2mix dataset~\cite{hershey2016deep,isik2016single}, which contains 30 hours of training and 10 hours of validation data. The mixtures are generated by randomly selecting 49 male and 51 female speakers and utterances in Wall Street Journal (WSJ0) training set si\_tr\_s, and mixing them at various signal-to-noise ratios (SNR) uniformly between 0 dB and 5 dB. 5 hours of evaluation set is generated in the same way, using utterances from 16 unseen speakers from si\_dt\_05 and si\_et\_05 in the WSJ0 dataset.

In this work, we shift the window around raw waveform by 5ms and produce a set of frames at 10ms intervals. Thus structure of the FurcaNet instance used in this work is as the following. The frontend GCNN has 5 1D GConv layers. Since the input to the FurcaNet is a speech frame of 10ms (80 sample points), thus the size of the first convolution kernel is 80, and the other 4 1D GConv layers are with kernel of size 1000. Behind each GConv layer, we add a layer normalization operation~\cite{ba2016layer} in order to stabilize the training. Then 2 BiLSTM layers with 1000 hidden units in each direction are employed after the GCNN. The DNN has 2 hidden layers of 2000 nodes each.

\subsection{Training trick}
\begin{figure}[htb]
\centering
\includegraphics[width=0.45\textwidth]{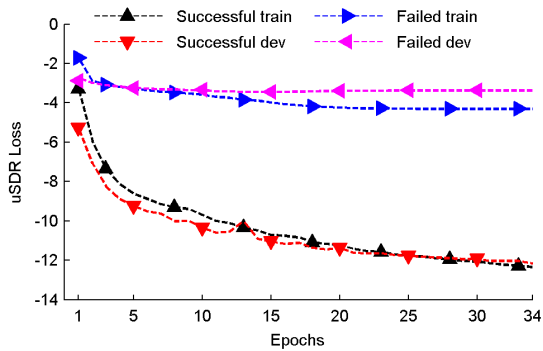}
\caption{The success and failure examples of uSDR loss optimization.}
\label{fig:local_min}
\end{figure}

During training Adam~\cite{kingma2014adam} serves as the optimizer to minimize the uSDR loss with initial learning rate of 0.001 and scale down by 0.5 when the training loss increased on the development set. Each mini-batch had 8 randomly selected utterances.
The uSDR loss function is a bit hard to optimize. Adam~\cite{kingma2014adam} often got stuck in a local minimum. As the Fig.~\ref{fig:local_min} shows, the horizontal axis is the trained epochs and the vertical axis is the negative SDR.
In failure training process, uSDR stuck at 3dB or 4dB, and do not go any further.
We found an ad-hoc trick to deal with this problem, since the weights of the FurcaNet is randomly initialized, we restart the training program directly until that the initial SDR (before any training epochs) on the development dataset is greater than a threshold (for example -30dB, the Fig.~\ref{fig:local_min} only shows the SDRs after each training epochs), then we will let the program start training.

\subsection{Results}

We evaluate the systems with the SDR improvement (SDRi)~\cite{fevotte2005bss,vincent2006performance} metrics used in~\cite{isik2016single,luo2018speaker,xu2018single,chen2017deep,kolbaek2017multitalker}.
The original SDR, that is the average SDR of mixed speech $y(t)$ for the original target speech $x_1(t)$ and $x_2(t)$ is 0.15.
Table~\ref{tab:sdri} lists the average SDRi obtained by FurcaNet and almost all the results in the past two years, where IRM means the ideal ratio mask
\begin{equation}
M_s=\frac{|X_s(t,f)|}{\sum_{s=1}^{S}|X_s(t,f)|}
\label{eq:irm}
\end{equation}
applied to the STFT $Y(t,f)$ of $y(t)$ to obtain the separated speech, which is evaluated to show the upper bounds of STFT based methods, where $X_s(t,f)$ is the STFT of $x_s(t)$.
In this experiment,
Chimera++~\cite{wang2018alternative,wang2018end} gives the best SDRi in all baselines shown in Table~\ref{tab:sdri}.
FurcaNet has achieved an improvement of 1.3dB SDRi compared with this best baseline
FurcaNet has achieved the most significant performance improvement compared with baseline systems, and it break through the upper bound of STFT based methods.

\begin{table}[th]
\caption[sdri]{SDRi (dB) in a comparative study of different separation methods on the WSJ0-2mix dataset. * indicates our reimplementation of the corresponding method.}\label{tab:sdri}
\centering
\begin{tabular}{|c|c|}
\hline
Method & SDRi  \\
\hline
DPCL~\cite{hershey2016deep} & 5.9  \\
\hline
DPCL* & 10.7  \\
\hline
DPCL++~\cite{isik2016single} & 10.8  \\
\hline
DANet~\cite{chen2017deep} & 10.5 \\
\hline
ADANet~\cite{luo2018speaker} & 10.5 \\
\hline
uPIT-BLSTM~\cite{yu2017permutation} & 10.0 \\
\hline
cuPIT-Grid-RD~\cite{xu2018single} & 10.2 \\
\hline
CBLDNN-GAT~\cite{li2018cbldnn} & 11.0 \\
\hline
TasNet~\cite{luo2017tasnet} & 11.2 \\
\hline
TasNet* & 11.8 \\
\hline
Chimera++~\cite{wang2018alternative,wang2018end} & 12.0 \\
\hline
FurcaNet & 13.3 \\
\hline
IRM & 12.7 \\
\hline
\end{tabular}
\end{table}

\section{CONCLUSION}
In this study, we proposed an end-to-end architecture called FurcaNet for monaural speech separation. FurcaNet can combine the advantages of different neural networks such as GCNN, BiLSTM, and DNN, and at the same time
it can directly optimize parameters using perceptual indicators such as SDR.
Our results on two-speaker mixed speech separation task indicate that FurcaNet can
achieve a state-of-the-art performance.
Future research would include extending the experiment to three-speaker mix task to see whether it is independent of the number of sound sources.

\section{ACKNOWLEDGMENT}
We would like to thank Jian Wu at Northwestern Polytechnical University, Yi Luo at Columbia University, and Zhong-Qiu Wang at Ohio State University for valuable discussions on WSJ0-2mix database, DPCL, and end-to-end speech separation.

\bibliographystyle{IEEEbib}
\bibliography{refs}

\end{document}